\documentclass[11pt] {article}
\usepackage{amsfonts}
\usepackage{amssymb}
\usepackage{graphicx}

\title{Testing a string dilaton model with experimental and
  observational data} 
\author{Susana J.Landau$^{1,2}$ \thanks{e-mail: {\tt slandau@df.uba.ar}} \and
  Melina Bersten$^{1,3}$  \thanks{e-mail: {\tt mbersten@astro.puc.cl  }} \and Pablo Sisterna
  \thanks{e-mail: {\tt sisterna@museodelmar.org}} \and Hector Vucetich$^{1}$
  \thanks{e-mail: {\tt vucetich@fcaglp.unlp.edu.ar}} \\  \small$^1$ Facultad
  de Ciencias Astron\'{o}micas y  
Geof\'{\i}sicas,\\ \small Universidad Nacional de La Plata\\ \small
  Paseo del Bosque 
S$/$N, CP 1900 La Plata, 
Argentina\\ \small $^2$ Departamento
de F\'{\i}sica, Facultad de Ciencias Exactas y Naturales,\\ 
\small Universidad Nacional de Buenos Aires\\ \small Pabell\'{o}n I,
  Ciudad 
  Universitaria, 
1428, Buenos Aires, Argentina \\
\small $^3$ Departamento de Astronomia y
Astrofisica \\ \small Pontificia Universidad Catolica de
Chile\\
 \small Av. Vicuña Makena 4860, 782-0436 Macul,
Santiago, Chile}

\begin{document}

\maketitle                 

\pagestyle{myheadings} 
\thispagestyle{plain}         
\markboth{Landau et al}{Testing a string dilaton model with
  experimental and observational data}  
\setcounter{page}{1}    

We test the prediction of the time variation of the fine structure
constant in the string dilaton model proposed by Damour and
Polyakov. First, we analize the dependence of all available
observational and experimental data with the fine structure constant
variation. Furthermore, we obtain the prediction of the time variation
of the fine structure constant including the renormalization group
correction. Finally, we use the data set to perform a statistical
analyisis. This analysis enables us to determine that the the dilaton
model is in agreement with most of the data. Finally, constraints on
the free parameters of this model are obtained.

\section{Introduction}

The attempt to unify all fundamental interactions resulted in the
development of multidimensional theories like string derived field
theories \cite{Wu86,Maeda88,Barr88,DP94,DPV2002a,DPV2002b}, related
brane-world theories \cite{Youm2001a,Youm2001b,branes03a,branes03b},
and (related or not) Kaluza-Klein theories
\cite{Kaluza,Klein,Weinberg83,GT85,OW97}.  Among these theories, there
are some in which the gauge coupling constants may vary over
cosmological time scales. On the other hand, a theoretical framework
based on first principles, was developed by Bekenstein
\cite{Bekenstein82} and later improved by Barrow, Sandvik and Magueijo
\cite{BSM02} in order to study the possible time variation of the fine
structure constant.  Furthermore, this model was generalized to study
the variation of the strong coupling constant by Chamoun et
al.\cite{CLV01}.

Different
versions of the theories mentioned above predict different time
behaviours of the gauge coupling constants. Thus, bounds obtained from
astronomical and geophysical data are an important tool to test the
validity of these theories.

The experimental research can be grouped into astronomical and local
methods. The latter ones include geophysical methods such as the
natural nuclear reactor that operated about $1.8\ 10^9$ years ago in
Oklo, Gabon \cite{DD96,Fujii00,Fujii02}, the analysis of natural
long-lived $\beta$ decayers in geological minerals and meteorites
\cite{Dyson67,SV90,Smolliar96} and laboratory measurements such as
comparisons of rates between clocks with different atomic number
\cite{PTM95,Sortais00,Marion03,Bize03,Fischer04,Peik04}. The
astronomical methods are based mainly in the analysis of spectra form
high-redshift quasar absorption systems
\cite{CS95,VPI96,Webb99,Webb01,Murphy01a,Murphy01b,LE02,Ivanchik02,%
Murphy03b,Ivanchik03,Bahcall04}.  Although, most of the previous
mentioned experimental data gave null results, evidence of time
variation of the fine structure constant was reported recently from
high-redshift quasar absorption systems
\cite{Webb99,Webb01,Murphy01a,Murphy01b,Murphy03b,Ivanchik03}. However,
other recent independent analysis of similar data
\cite{Bahcall04,MVB04,QRL04,Srianand04} found no variation. On the
other hand, measurements of molecular hydrogen \cite{Ivanchik02,
Ivanchik03} reported a variation of the proton to electron mass $\mu =
\frac{m_p}{m_e}$.  Furthermore, the time variation of the gauge
coupling constants in the early universe can be constrained using data
from the Cosmic Microwave Background (CMB)
\cite{BCW01,AV00,Martins02,Rocha03} and the primordial abundances of
light elements \cite{Iguri99,Ichi02,Nollet}.

Damour and Polyakov \cite{DP94} proposed a string dilaton model in
which the variation of the gauge coupling constants is driven by the
time evolution of the dilaton field. In this model, the dilaton
remains massles and in consequence all coupling constants and masses
of elementary particles are time dependent. In this chapter we limit
ourselves to the variation of the fine structure constant. We compare
the prediction of the dilaton model with all available data on time
variation of the fine structure constant but the nucleosynthesis
data. In section \ref{bounds} we describe carefully the data set
considered. In section \ref{theorie} we briefly review the Damour and
Polyakov proposal \cite{DP94} and obtain the analytic expression for
the prediction of the time variation of the fine structure constant
including the renormalization group corrections. Finally, in section
\ref{results} we present the results of the statistical analysis and
briefly discuss our conclusions.

\section{Bounds from astronomical and geophysical data}
\label{bounds}

In this section, we review all availabe bounds on time variation of
the fine structure constant. We discuss the relation between the
observable quantities and the variation of $\alpha$ in each case. We
also describe very carefully the error we consider for each data.

\subsection{The Oklo Phenomenon}

One of the most stringent limits on time variation of the fine
structure constant $\alpha$ follows from an analysis of isotope ratios
in the natural uranium fission reactor that operated $1.8\times 10^9$
yrs ago at the present day site of the Oklo mine in Gabon,
Africa. From an analysis of nuclear and geochemical data, the
operating conditions of the reactor could be reconstructed and the
thermal neutron capture cross sections of several nuclear species
measured. In particular, a shift in the lowest lying resonance level
in $^{149}{\rm Sm}: \Delta = E_r^{149{\rm(Oklo)}} -
E_r^{149{\rm(now)}}$ can be derived from a shift in the neutron
capture cross section of the same nucleus \cite{Fujii00,DD96}.  The
first estimate of a change in the resonance energy was peformed by
Damour and Dyson \cite{DD96}. This bound was re-examined by Fujii et
al \cite{Fujii00} using new samples of $^{149}{\rm Sm}$, $^{155}{\rm
Gd}$ and $^{157}{\rm Gd}$. In their analysis they take the effect of
contamination into account, assuming the same contamination parameter
for all samples. They obtain the following bound:
\begin{equation}
\Delta = E_r^{Oklo} - E_r^{today} = \left(9 \pm 11\right) \, 10^{-3} eV
\end{equation}
The shift in $\Delta $ can be translated \cite{DD96} into a bound on a
possible difference between the value of $\alpha $ during
the Oklo phenomenon and its value now, as follows:
\begin{equation}
\Delta =\alpha \frac{\partial E_r}{\partial \alpha }\frac{\Delta \alpha }
\alpha 
\end{equation}
where $ \frac{\partial E_r}{\partial \alpha } = 10^6$. The dependence
of the resonance energy with other fundamental constants has been
analized by Damour and Dyson \cite{DD96} and Sisterna and Vucetich
\cite{SV91}.

\subsection{Long-lived $\beta $ decayers}

The half-life of long-lived $\beta $ decayers has been determined
either in laboratory measurements or by comparison with the age of
meteorites, as found from $\alpha $ decay radioactivity analysis.  In
table \ref{decaimientos}, we show $\frac{\Delta \lambda}{\lambda}$ for
three diferent decayers: $^{187}\rm{Re}, ^{40}\rm{K}, ^{87}\rm{Rb}$.
Sisterna and Vucetich \cite{SV90} have derived a relation between the
shift in the half-life of long lived $\beta $ decayers and a possible
variation between the values of the fundamental constants $\alpha
,\Lambda _{QCD}$ and $G_F$ at the age of the meteorites and their
value now. In this chapter, we only consider $\alpha$ variation and
therefore, the following equation holds:
\begin{equation}
\frac{\Delta \lambda}{\lambda} = a \frac{\Delta \alpha}{\alpha}
\end{equation}
where $a=21600, 46, 1070$ for $^{187}\rm{Re}, ^{40}\rm{K},
^{87}\rm{Rb}$ respectively \cite{SV91}.

\begin{table}[h]
\label{decaimientos}
\caption{The table shows the $\beta$ decayer, the difference of half life measured in the laboratory and from meteorites, the corresponding error and reference.}
\begin{center}
\begin{tabular}{cccc}
\hline
\hline
  $\beta$ decayer & $\frac{\Delta \lambda}{\lambda}$ & $\sigma\left(\frac{\Delta \lambda}{\lambda}\right)$ &  Reference \\
\hline
$^{187}Re$ & $-1.6 \, 10^{-2}$ & $1.6 \, 10^{-2}$ & \cite{Smolliar96}\\
$^{40}K$ & $0$  & $1.3 \, 10^{-2}$  & \cite{Dyson67}\\
$^{87}RB$ & $0$ &  $1.3 \, 10^{-2}$  & \cite{Dyson67}\\
\hline
\hline
\end{tabular}
\end{center}
\end{table}

\subsection{Atomic clocks}

The comparison of different atomic transition frequencies over time
can be used to determine the present value of the temporal derivative
of $\alpha$. Indeed, the more stringent limits on the variation of
$\alpha$ are obtained using this method.

Hyperfine transition frequencies have the following dependence with $\alpha$:
\begin{equation} 
\nu_{Hyp} \sim \alpha^2 \frac{\mu}{\mu_N}\frac{m_e}{m_p} R_{\infty} c
F_{REL}(\alpha Z) 
\end{equation}
where $\mu$ is the nuclear magnetic moment, $\mu_B$ is Bohr magneton,
$R_{\infty}$ is Rydberg's constant, $m_p$ and $m_e$ are the proton and
electron mass and $F_{REL}$ is the relativistic contribution to the
energy. In such way, the comparison of rates between clocks based on
hyperfine transitions in alkali atoms with different atomic number $Z$
can be used to set bounds on $\alpha^k \frac{\mu_{A_1}}{\mu_{A_2}}$
where $k$ depends on the frequencies measured and $\mu_{A_i}$ refers
to the nuclear magnetic moment of each atom. As explained above, we
are considering only $\alpha$ variation, but it is important to keep
in mind that this type of experiments actually constrain a combination
of $\alpha$ and other fundamental quantities. The first three entries
of table \ref{atomic} show the bounds on $\frac{\Delta
\alpha}{\alpha}$ obtained comparing hyperfine transition frequencies
in alkali atoms.  On the other hand, an optical transition frequency
has a different dependence on $\alpha$:
\begin{equation}   
\nu_{opt} \sim R_{\infty} B F_i(\alpha)
\end{equation}
where $B$ is a numerical constant assumed not to vary in time and
$F_i(\alpha)$ is a dimensionless function of $\alpha$ that takes into
account level shifts due to relativistic effects. Thus, comparing an
optical transition frequency with an hiperfine transition frequency
can be used to set bound on $\alpha^k \frac{m_e}{m_p}
\frac{\mu_A}{\mu_B}$. Again, we will only consider $\alpha$
variation. Different authors \cite{Bize03,Fischer04,Peik04} have
measured different optical transitions and set bounds on the variation
of $\alpha$ using different methods. Fischer et al \cite{Fischer04}
have considered the joint variation of $\alpha$ and
$\frac{\mu_{Cs}}{\mu_B}$. We have reanalized the data of
ref. \cite{Fischer04}, considering only $\alpha$ variation, yieding
the fifth entrie of table \ref{atomic}.  On the other hand, Peik et
al, have measured an optical transition frequency in $^{171}Yb+$ with
a cesium atomic clock. Furthermore, they perform a linear regression
analysis using this result toghether with other optical transition
frequency measurements \cite{Bize03,Fischer04}. On one hand, a linear
regression analysis with three points has no statistical
significance. On the other hand, we have already considered the other
data. Therefore, we have also reanalized the data, using only the
comparison between $Yb+$ and $Cs$ frequency, yielding the sixth entrie
in table \ref{atomic}.

\begin{table}[tbp]
\label{atomic}
\caption{The table shows the clocks compared, the value of
  $\frac{\Delta \alpha}{\alpha}$ and its corresponding error, the time
  interval for which the variation was measured and the reference.} 
\begin{center}
\begin{tabular}{ccccc}
\hline
\hline
 Frequencies  & $\frac{\Delta\alpha}{\alpha}$ &
 $\sigma\left(\frac{\Delta \alpha}{\alpha}\right)$   & $\Delta t$(yr)
 &  Reference \\ 
\hline
Hg+ and   H maser & $0$ & $1.4 \, 10^{-14}$ & 0.38  &  \cite{PTM95}\\
 Cs and Rb & $8.4 \, 10^{-15} $  & $13.8 \, 10^{-15}$ & 2 & \cite{Sortais00}\\
 Cs and Rb & $-2 \, 10^{-16} $  & $8 \, 10^{-15}$ & 5 & \cite{Marion03} \\
Hg and Cs & $ 0$ & $2.4 \, 10^{-15}$ & 2 & \cite{Bize03} \\
H and Cs & $5.7 \, 10^{-15}$ & $11.2 \, 10^{-15}$ & 5 & \cite{Fischer04} \\
Yb and  Cs & $-1.6 \, 10^{-15}$ & $5.9 \, 10^{-15}$ & 2.8 & \cite{Peik04} \\
\hline
\hline
\end{tabular}
\end{center}
\end{table}

\subsection{Quasar absorption systems}

Quasar absorption systems present ideal laboratories  to
search for any temporal variation in the fundamental constants. The
continuum spectrum of a quasar was formed at an epoch corresponding to
the redshift $z$ of main emission details specified by the
relationship:
\begin{equation}
\lambda _{obs}=\lambda _{lab}\left( 1+z\right) .
\end{equation}

Quasar spectra of high redshift show the absorption resonance lines of
 alkaline ions like CIV, MgII, FeII, SiIV and others, corresponding
to the $ S_{1/2}\rightarrow P_{3/2}\left( \lambda _1\right) $ and
$S_{1/2}\rightarrow P_{1/2}\left( \lambda _2\right) $ transitions. The
relative magnitude of the fine splitting of the corresponding
resonance lines is proportional to the square of the fine structure
constant $\alpha $ to lowest order in $\alpha $.
\begin{equation}
\frac{\Delta \lambda }\lambda =\frac{\lambda _1-\lambda _2}\lambda \sim
\alpha ^2
\end{equation}

Therefore, any change in the value of $\alpha $ at redshift $z$ with
respecto to the laboratory value, can be measured from the separation
of the doublets $ \Delta \lambda $ as follows:
\[
\frac{\Delta \alpha }\alpha =%
{\textstyle {1 \over 2}}c_r
\left[ \frac{\left( \frac{\Delta \lambda }\lambda \right) _z}{\left( \frac{%
\Delta \lambda }\lambda \right) _{now}}-1\right] 
\]
where $c_r$ is a correction term which depends on the ion
considered. This method is known in the literature as the Alkali
Doublet (AD) method.  Several authors have applied this method to SiIV
doublet absorption lines systems at different redshifts ($1 < z <
3.6$).  We show the average values they obtain in table
\ref{dobletesalcalinos2}. On the other hand, in order to perform our
statistical analysis, we consider the individual data and
corresponding errors for each absorption cloud which can be found in
the references.

Bahcall et al \cite{Bahcall04} use strong nebular emission lines of O
III to constrain the variation of $\alpha$. Again, in this case,
measuring the relative separation of the doublet give a constraint on
$\alpha^2$ at a given redshift. Assuming a linear variation of
$\alpha$ with time, the authors find $\frac{\Delta \alpha}{\alpha} =
(-0.7 \pm 1.4) \times 10^{-4}$.  In this work , we want to test a
theoretical model and thus the data must be model independent. Thus,
we will consider the individual measurements on the variation of
$\alpha$ belonging to the standard sample consisting in 42 quasar
absorption systems, which are listed in table 4 of reference
\cite{Bahcall04}.

\begin{table}[tbp]
\caption{The table shows the redshift interval, the average value and
  standard deviation of  $\frac{\Delta \alpha}{\alpha}$ in units of
  $10^{-4}$b obtained using the AD method  and the corresponding
  reference.} 
\label{dobletesalcalinos2}
\begin{center}
\begin{tabular}{cccc}
\hline
\hline
 $z_{abs}$ & $\frac{\Delta \alpha}{\alpha}$ &  $\sigma \left(
\frac{\Delta \alpha}{\alpha} \right) $ & Reference \\ 
 &   $\times 10^{-4}$ & $\times 10^{-4}$  &  \\
\hline
$2.6 < z < 3.6 $ &  $-0.35$ & $3.5$  & \cite{CS95} \\
$2.8 < z < 3.05 $ &  $0.21$ & $1.4$  & \cite{VPI96}\\
$2 < z < 3  $ &  $-0.05$ & $0.13$  & \cite{Murphy01b}\\
$1.18 < z < 1.83 $& $-0.31$ & $0.85$ & \cite{MVB04} \\
\hline
\hline
\end{tabular}
\end{center}
\end{table}

An improvement to the AD Method was proposed by Webb et al
\cite{Webb99,Dzuba99}. The new method called in the literature the
Many Multiplet (MM) method compares transitions of different species,
with widely differing atomic masses toghether with different
transitions of the same species.  If we consider a many electron atom
or ion, the relativistic correction to the energy of the external
electron can be written as:

\begin{equation}
\Delta \sim {\left(Z_n \alpha\right)}^2 {|E|}^{3/2} \left[\frac{1}{j +
    1/2} - C\left(j,l\right)\right] 
\label{er}  
\end{equation}
where $Z_n$ is the nuclear charge, $E$ is the electron's energy, $j$
and  $l$ are the electrons's total and orbital angular momentum
respectively. Moreover, $C\left(j,l\right)$ is the contribution added
due to the many body effects. Again, as in the case of the atomic
clocks, the relativistic contribution is proportional de $Z \alpha^2$
and therefore comparing different transitions of the same atom or
transitions of different atoms, is a useful tool to put bounds on
$\alpha$. Moreover, the energy equation for a transition to the ground
state, within a particular multiplet, at a redshift $z$ reads: 
\begin{equation}
E_z = E_c + \left[Q_1 + K_1 \left(\vec L . \vec S\right)\right] Z_n^2 \left[{\left(\frac{\alpha_z}{\alpha_0}\right)}^ 2 - 1\right] + K_2 \left(\vec L . \vec S\right) Z_n^4 \left[{\left(\frac{\alpha_z}{\alpha_0}\right)}^ 4 - 1\right] 
\label{MM}
\end{equation}
where $\vec L$ and $\vec S$ are the electron total orbital angular and
spin momentum respectively, $E_c$ is the energy of the configuration
centre, $Q_1$, $K_1$ y $K_2$ are relativistic coefficients which have
been accurately computed by Dzuba et
al. \cite{Dzuba99,Dzuba99b,Dzuba01}. The limits over the fine
structure constant variation are obtained fitting Voigt profiles to
the absorption features in several different transitions. To the three
usual fit parameters : column density, Doppler width and redshift,
they add $\frac{\Delta \alpha}{\alpha}$. In such way, this method
makes possible to gain an order of order of magnitude in sensibility
with respect to the AD method. On the other hand, the simultaneous fit
of the Voigt profiles, can allow that unknown systematic errors hide
under the variation of $\alpha$.  As mentioned before, this method
provides the only results consistent with a time varying fine
structure constant \cite{Webb99,Webb01,Murphy01a,Murphy03b}. In a
recent work \cite{Murphy03b} they estimate $\frac{\Delta
\alpha}{\alpha}= (-0.543 \pm 0.116) \times 10^{-5}$ for 128 absorption
systems over the redshift range $0.2 < z < 3.7 $ confirming their
previous results. For our analysis we consider the individual values
obtained for each absorption system, listed in table 3 of reference
\cite{Murphy03b}. As suggested by the authors we add to the each
individual error listed in table 3 of ref. \cite{Murphy03b}, an
additional random error of $2.09 \times 10^{-5}$. 

On the other hand,
the same method was applied by other authors
\cite{Srianand04,Chand04}, who also use a stringent selection criteria
of the samples, discarding weak and blended lines among others. They
have obtained no variation of $\alpha$ for a high quality quasar
spectra obtained using VLT over the redshift range $0.4 < z <
2.3$. Again, in this case, we use the individual estimates obtained
for each redshift listed in table 3 of reference \cite{Chand04}.  On
the other hand, the standard MM tecnique can be revised to avoid the
deficiencies pointed out earlier and in the literature
\cite{Bahcall04,Le03,QRL04}. In fact, for $\frac{\Delta
\alpha}{\alpha} <<1$ equation \ref{MM} can be re-written as follows:
\begin{equation}
z_i = z_{\alpha} + \kappa_{\alpha} Q_i 
\end{equation}
where $z_i$ denotes the observed redshift and $Q_i$ the sensitivity
coefficient corresponding to the lines $i$, and the slope parameter
$\kappa_{\alpha}$ is given by:
\begin{equation}
\kappa_{\alpha} = -2 \left( 1 + z_{\alpha} \right) \frac{\Delta \alpha}{\alpha}
\end{equation}

In such way, it is possible to estimate the variation of $\alpha$ from
linear regression analysis of the position of the line centroids in an
absorption component. The accuracy of the regression analysis will be
improved, if several absorption line samples are combined. This
improved method called in the literature as Revised Many Multiplet
(RMM) method, was applied by Levshakov \cite{Le03} and Quast et al
\cite{QRL04} to a homogeneous sample of FeII lines at redshift $z =
1.149$ and $z=1.15$ to obtain $\frac{\Delta \alpha}{\alpha}= (1.1 \pm
1.1) \times 10^{-5}$ and $\frac{\Delta \alpha}{\alpha}= (-0.4 \pm 4.6)
\times 10^{-5}$ respectively.

On the other hand, $OH$ lines can provide precise constraints on
cosmic evolution of $\alpha$, the proton $g$ factor and the ratio of
electron to proton mass. Again, in this chapter, we will only consider
$\alpha$ variation. Darling \cite{Darling04} reported the detection of
of the satellite $18$ cm $OH$ conjugate lines at $1612$ and $1720$ MHz
in the $z=0.2467$ molecular absorption system toward the radio source
PKS $1413+135$. Conjugate lines profiles guarantee that both lines
originate in the same molecular gas. On the other hand, the 18 cm $OH$
lines can be decomposed into a $\Lambda$-doubled term chich depends
weakly on $\alpha$ and a hiperfine term which has a strong $\alpha^4$
dependence. From these, sums and differences of lines can form pure
$\Lambda$-doubled and pure hyperfine quantities:
\begin{center}
\begin{eqnarray}
\Sigma \nu = \nu_{1720} + \nu_{1612}= 2 \Lambda \alpha^{0.4} \nonumber\\ 
\Delta \nu = \nu_{1720} - \nu_{1612}= 2 \left(\Delta^+ +
\Delta^-\right) \alpha^4 \nonumber  
\end{eqnarray}
\end{center}
where $\Delta^+ = 9.720$ MHz and $\Delta^- = 9.375$ MHz and $\Lambda =
11926 $MHz .  In such way, Darling obtains $\frac{\Delta
\alpha}{\alpha} = \left(0.5 \pm 1.3 \right) \times 10^{-5}$.

Moreover, the ratio of the hyperfine 21 cm absorption
transition of neutral hydrogen $\nu _a$ to an optical resonance
transition $ \nu _b$ is proportional to $x=\alpha ^2g_p\frac{me}{mp}$
where $g_p$ is the proton $g$ factor. Thus, a
change of this quantity will result in a difference in the redshift
measured from 21 cm and optical absorption lines as follows:
\begin{equation}
\frac{\Delta x}x=\frac{z_{opt}-z_{21}}{\left( 1+z\right) }
\end{equation}
 
So, combining the measurements of optical and radio redshift, a bound
on $\alpha^2$ can be obtained. Table \ref{opticoradio} shows the
bounds obtained by different authors using this method. This method
has the inconvenience that it is difficult to determine if both radio
and optical lines originate at the same absorption system. Thus, a
difference in the velocity of the absorption clouds could hide in a
variation of $\alpha$.

 \begin{table}[tbp]
\caption{The table shows the absorption redshift, the value and
  standard deviation  of $\frac{\Delta \alpha}{\alpha}$ obtained
  comparing optical and radio lines in units of $10^{-4}$ and the
  corresponding references} 
\label{opticoradio}
\begin{center}
\begin{tabular}{cccc}
\hline
\hline
 $z_{abs}$ & $\frac{\Delta \alpha}{\alpha}$ &  $\sigma \left(
\frac{\Delta \alpha}{\alpha} \right) $ & Reference \\ 
\hline
\hline
$1.77 $  &  $-0.035$ & $0.055$  & \cite{CS95} \\
$0.52$  &  $0$ & $0.6$  &  \cite{WBR76}\\
$0.69$  &  $0$ & $1.4$  & \cite{SM79}\\
\hline
\hline
\end{tabular}
\end{center}
\end{table}

The ratio of the rotational transition frequencies of
diatomic molecules such as CO to the 21 cm hyperfine transition in
hydrogen is proportional to $y=g_p\alpha ^2$. Thus, any variation in
$y$ would  be observed as a difference in the 
redshifts measured from 21 cm and molecular transition lines:
\begin{equation}
\frac{\Delta y}y=\frac{z_{mol}-z_{21}}{\left( 1+z\right) }
\end{equation}

Murphy et al. \cite{Murphy02} have placed upper limits at redshift $
z=0.25$, $\frac{\Delta \alpha}\alpha =\left(-0.1 \pm 0.22\right)
\times 10^{-5}$ and at redshift $z=0.68$, $\frac{\Delta \alpha}\alpha
=\left(-0.08 \pm 0.27\right) \times 10^{-5}$.

\subsection{Cosmic Microwave Background}

Any variation of the fine structure constant $\alpha$ alters the
physical conditions at recombination and therefore changes the cosmic
microwave background (CMB) fluctuation spectrum. The dominant effect
is a change in the redshift of recombination, due to a shift in the
energy levels and, in particular the binding energy of Hidrogen. The
Thompson scattering cross section is also changed for all particles,
being proportinal to $\alpha^2$. A different value of $\alpha$ at
recombination affects the CMB fluctuation spectrum in two ways: i) a
shift in the Doppler peaks position and ii) a change in the amplitude
of the Doppler peaks.  On the other hand, the CMB fluctuacion spectrum
is sensitive to many cosmological parameters such as the density of
barionic and dark matter, the Hubble constant and the index of
primordial spectral fluctuations. Moreover, the effect of changing
this parameters is similar to a change in $\alpha$. Even though the
bounds obtained using this method are not very stringent, it is
important to consider a bound of $\alpha$ in the early universe.

Martins et al \cite{Martins02,Rocha03} have performed an estimation of
the fine structure constant toghether with other cosmological
parameters with the first year data of the Wilkinson Anisotropy Probe
(WMAP) \cite{wmapdata}. They established the following bound:
\begin{equation}
\frac{\Delta \alpha}{\alpha} =0.025 \pm 0.035  
\end{equation}

\section{Theoretical Model}
\label{theorie}

In this section, we review the theoretical model proposed by Damour
and Polyakov. We derive the expression for the observable quantity
$\frac{ \Delta \alpha}{\alpha}$, including the renormalization group
correction which was not considered in the original paper.

\subsection{The Gravity-dilaton-matter Model}

The existence of a massless scalar field coupled to gravity and matter
presents a host of effects that contradicts several
experiments. Universal couplings violate the strong equivalence
principle while non universal ones violate the universality of free
fall for different composite probes. The original proposal of ref
\cite{DP94} was a mechanism by which the dilaton, while still
massless, is atracted towards some maximum of its coupling functions
to the other fields (denoted as $B_a$). It is known that string loop
effects associated with worldsheets of arbitrary genus in intermediate
string sates can make $B_a=B_a(\Phi)$ to depend on $\Phi$ in a
non-monotonic way.  If the extrema of this functions differ from each
other, then the dilaton can still at present be sensitively changing
in time, producing unacceptable time dependences of fundamental
constants and violation to the weak equivalence principle. Conversely,
if the coupling is universal, then the extremum will be universal as
well and the cosmological evolution at all stages will conspire to
drive the dilaton almost precisely towards such a value, leaving a
very tiny remanent time dependence and very small violations of
universality of free fall.

Let us briefly describe the model. In the string frame metric denoted as 
$\hat g_{\mu\nu}$ the dilaton couples in a universal multiplicative way to 
all other fields at the string tree level. However, when taking the full 
string-loop expansion, the effective action takes the general form: 
\begin{eqnarray}
S &=&  \int d^4x \sqrt{\hat g}\left(\frac{B_g(\Phi)}{\alpha^{\prime}}\hat 
R +
\frac{B_{\Phi}(\Phi)}{\alpha^{\prime}}[4 \Box  \Phi-4(\hat \nabla 
\Phi)^2]\right. \nonumber 
\\
&&\left.-B_F(\Phi)\frac{k}{4}\hat F^2-B_{\psi}(\Phi)\bar{\hat \psi}
\hat D \hat  
\Psi-i \bar{\hat \Psi} m_f \hat \Psi \right) 
\label{accion}
\end{eqnarray} 
where $\hat R$ is the Ricci scalar corresponding to the string metric, 
$\Phi$ the dilaton, $\hat F^2= F^a_{\mu \nu}F^{a \mu \nu}$ with
$ F^a_{\mu\nu} =\partial_\mu( A^a_\nu)-\partial_\nu(A^a_\mu)+
f^{abc}A^b_\mu A^c_\nu $ includes all the gauge fields, and $\hat D$ is the 
gauge covariant derivative.

Following ref.\cite{DP94} we introduce the Einstein metric 
$g_{\mu \nu}= C B_g(\Phi) \hat g_{\mu \nu}$ 
and a convenient $\Phi$-dependent rescaling:
\begin{equation}
\phi = \int  d\Phi \left(\frac{3}{4}(\frac{B'_g}{B_g})^2 + 2  
\frac{B'_\Phi}{B_g}+2\frac{B_\Phi}{B_g} \right)^{\frac{1}{2}}
\end{equation}
where $B'_g=dB_g/d\Phi$ and $C$ is a constant such as $C B_g(\Phi_0)=1$ 
today.
Also we rescale the Dirac fields:
\begin{equation}
\Psi= C^{-\frac{3}{4}} B_g^{-\frac{3}{4}} B_\Psi^{\frac{1}{2}}\hat \Psi 
\end{equation}
finally obtaining the action, decomposed into a gravitational sector 
$(g_{\mu\nu},\phi)$ and a matter sector $(\psi, A, \ldots)$:
 
\begin{equation}
S[g,\Phi,\Psi,A,...]= S_{g,\Phi}+S_m
\end{equation}

\begin{equation}
S_{g,\Phi}= \int d^4x \sqrt{\hat g}\left(\frac{R}{4q} -
\frac{(\nabla \Phi)^2}{2q}\right)
\end{equation}

\begin{equation}
S_m= \int d^4x \sqrt{\hat g}\left(-\bar{\hat \psi} \hat D \hat \Psi
- \frac{k}{4} B_F(\Phi)\hat F^2-i \bar{\hat \Psi} m_f \hat \Psi\right) 
\label{sm}
\end{equation}
where $q=4\pi G$ and $G$ is Newton's constant (the action does not include 
de poorly-known Higgs sector). We see that the unified gauge coupling constant 
is given by:
\begin{equation}
g^{-2} = \alpha_{GUT}^{-1}= k B_F(\Phi), \label{alphagut}
\end{equation}
a function of the dilaton. Rescaling the metric means rescaling inversely 
all energies, so 
the string cutoff mass scale becomes dilaton dependent in Einstein units:
\begin{equation}
\Lambda_s(\phi) = C^{-1/2} B_g^{-1/2}(\phi) \hat\Lambda_s.
\end{equation}
We still need to land on observational energy scales. In the case of an 
asymptotically free theory (e.g. QCD), at the one loop level the infrared 
confinement mass scale $\Lambda_{conf}$ is related to the cutoff scale as
\begin{equation}
\Lambda_{conf}\propto\Lambda_s \exp (-8\pi^2 b^{-1} g^{-2}) = 
C^{-1/2} B_g^{-1/2}(\phi) \exp [-8\pi^2 b^{-1} k B_F(\phi)] \hat\Lambda_s, 
\label{lambdaexp}
\end{equation}
with $b$ dependent on the particular gauge and matter fields.

It is assumed that the mass of any particle $A$ depends in a
non-trivial way on the VEV of the dilaton through the $B$'s:
\begin{equation}
m_A(\phi)= m_A[B_g(\phi),B_F(\phi),...]
\end{equation}
the key hypothesis of ref.\cite{DP94} being that the functions
$m_A(\phi)$ all have the same minimun $\phi_m$ that monotonically
drives the dilaton to that value, as mentioned above.

Another key assumption, inspired by Eq.\ref{lambdaexp} and the chiral limit 
in QCD (all energies proportional to $\Lambda_{QCD}$), is that any mass 
(composite or not) have the following form:
\begin{equation}
m_A(\phi) = \mu_A B^{-1/2}(\phi) \exp [-8\pi^2 \nu_A B(\phi)] \hat\Lambda_s, 
\label{massexp}
\end{equation}
where it has been assumed that all couplings $B_i$ depend on a common
function $B$ through functions $f_i$ (sufficient condition to have a
common extremum); $\nu_A$ and $\mu_A$ are pure numbers of order unity.
From here we see that a minimum of $m_A(\phi)$ corresponds to a
maximum of $B(\phi)$, and hence a minimum of $\ln
B^{-1}(\phi)$. Following \cite{DP94} we expand this function up to
second order around the minimum:
\begin{equation}
\ln B^{-1}(\phi) = \ln B^{-1}(\phi_m) + \frac{1}{2} \kappa (\phi-\phi_m)^2.
\label{lnB}
\end{equation}

From Eq.\ref{alphagut} we see that $\ln\alpha_{GUT}\propto\ln B^{-1}$. 
In the next section we will study the cosmological model rendering the 
time dependence of $\phi$. Now we focus on the relation between 
$\alpha_{GUT}$ and $\alpha$. First it is convenient to work at the 
unification scale $E_{GUT}$. $\alpha$ can be written in terms of the 
$U(1)\bigotimes SU(2)$ coupling constants as:
\begin{equation}
\frac{1}{\alpha(E_{GUT})}=\frac{5}{2}\frac{1}{\alpha_1(E_{GUT})}
+ \frac{1}{\alpha_2(E_{GUT})}=\frac{7}{2}\frac{1}{\alpha_{GUT}}
\end{equation}
where we have used that $\alpha_1(GUT)=\alpha_2(GUT)=\alpha_{GUT}$, the 
unification value. 

The renormalization group equation for the observable $\alpha$ is:
\begin{equation}
\frac{1}{\alpha(E_O)}=\frac{7}{2}\frac{1}{\alpha_{GUT}}+ \hat b \ln 
\left(\frac{E_{GUT}}{E_O}\right) \label{renor}
\end{equation}
where $\hat b$ depends on the unification model, and $E_O$ is the energy at 
which we measure $\alpha$. Given that we expect 
$\Delta \alpha/\alpha\leq 10^{-5}$, we can write:
\begin{equation}
\frac{\Delta \alpha}{\alpha} = \alpha 
\left(\frac{1}{\alpha_0}-\frac{1}{\alpha}\right) \simeq \alpha_0 
\left(\frac{1}{\alpha_0}-\frac{1}{\alpha}\right).
\end{equation}

From Eq.\ref{lambdaexp}, and assuming that $E_{GUT}\simeq \Lambda_s$,
we expect that:
\begin{equation}
\frac{E_{GUT}}{E_O} = C^{1/2} B_g^{1/2}(\phi) \exp [8\pi^2 b^{-1} k B_F(\phi)].
\end{equation}
Consequently we have:
\begin{equation}
\Delta\ln\left(\frac{E_{GUT}}{E_O}\right) = 
\frac{1}{2} \Delta\ln B_g + 8\pi^2 b^{-1} k \Delta B_F(\phi).
\end{equation}
Combining the last three equations we obtain:
\begin{equation}
\frac{\Delta \alpha}{\alpha}(t) = 
\left(\frac{7}{2} + 8\pi^2 \frac{\hat b}{b}\right) 
\alpha_0\Delta\alpha_{GUT}^{-1} + \frac{\hat b}{2}\alpha_0\frac{\Delta 
B_g}{B_g},
\end{equation}
where $\alpha_0=1/137$ is the present low energy value of the fine
structure constant.  Under the hypothesis of universality, we have
$B_g(\phi)=B_F(\phi)=B(\phi)$, and the last equation simplifies to
\begin{equation}
\frac{\Delta \alpha}{\alpha}(t) = 
\left(\frac{7}{2} + 8\pi^2 \frac{\hat b}{b} + \frac{\hat b}{2}\right) 
\alpha_0\frac{\Delta B}{B}={\cal A} \Delta\ln B^{-1}.\label{varalpha}
\end{equation}
We can estimate $\hat b$ from Eq.\ref{renor} 
($\alpha_{GUT}(t_0)\simeq \frac{1}{50}$). Regarding the 
string one-loop coefficient $b$ which relates differente mass scales, 
it depends on group parameters of order one (aswell as $\hat b$) and is 
positive. As we are seeking upper bounds for the parameter $\Delta\phi$ 
from the observational upper bounds on $\Delta\alpha$, we will use the 
conservative lower bound ${\cal A}=7/2+\hat b/2$.
From Eq.\ref{lnB} we can then write:
\begin{equation}
\frac{\Delta \alpha}{\alpha}(t) = 
\frac{1}{2}{\cal
  A}\kappa\Delta\left\{(\phi(t)-\phi_m)^2\right\}. \label{delta1} 
\end{equation}
In order to proceed we must solve the cosmological time evolution of the 
dilaton, which we do in the next section.

\subsection{Cosmological Solution}

We need to know the time evolution of the dilaton, which will be apparent 
after solving Einstein equations together with the dilaton equation. 
The equation derived from the action $S = S_{g,\Phi}+S_m$ are:
\begin{equation}
R_{\mu \nu}= 2 \partial_\mu \phi \partial_\nu \phi+2q 
(T_{\mu\nu}-\frac{1}{2}
g_{\mu \nu})
\label{campo1}
\end{equation}

\begin{equation}
\Box \phi=-q \sigma
\label{campo2}
\end{equation}
where
\begin{equation}
T^{\mu \nu}= 2 g^{-1/2}\frac{\delta S_m}{\delta g _{\mu \nu}}
\end{equation}

\begin{equation}
\sigma= 2 g^{-1/2}\frac{\delta S_m}{\delta \phi}
\label{sigma1}
\end{equation}
are the sources.

The several gasses (labeled by $A$) that describe the material content of the 
Universe, which we supposed weakely interacting, can be written as:
\begin{equation}
T^{\mu \nu}= \frac{1}{\sqrt{g(x)}}\sum_A \int ds_A m_A[\phi(x_a)] 
u^{\mu}_A u^{\nu}_A \delta^{(4)}(x-x_A)
\end{equation}
and consequently
\begin{eqnarray}
\sigma(x)&=& -\frac{1}{\sqrt{g(x)}}\sum_A \int ds_A \alpha_A[\phi(x_a)] 
m_A[\phi(x_a)]\delta^{(4)}(x-x_A) \nonumber \\ 
&=&\sum_A  \alpha_A[\phi(x_A)] T_A(x)
\label{sigma2}
\end{eqnarray} \\
where
\begin{equation}
\alpha_A(\phi)=\frac{\delta\ln{m_A(\phi)}}{\delta \phi}
\end{equation}
is a measure of the strenght of the dilaton coupling to the $A$ particles, 
$u^{\mu}_A = dx^{\mu}_A / ds_A$ and $T_A=-\rho_A+3P_A$ is the $A$ particles 
contribution to the trace of the energy-momentum tensor.

We consider the Friedmann cosmological model, in which
$ds^2=-dt^2+a(t)^2 dl^2$ where 
$dl^2=(1-Kr^2)^{-1} dr^2 +r^2(d\theta^2+\sin{\theta}^2 d\phi^2)$
with $K=0,1,-1$. Thus the gravitational and dilaton field equations 
simplify to:
\begin{displaymath}
-3\frac{\dot{a}}{a}=q(\rho+3P)+2\dot{\phi^2}
\end{displaymath}
\begin{equation}
\label{dil}
3H^2+3\frac{K}{a^2}=2q\rho+ \dot{\phi^2} \label{ecmov}
\end{equation}
\begin{displaymath}
\ddot{\phi}+3H\dot{\phi}=q\sigma
\end{displaymath}
where the dot denotes the time derivative $(d/dt)$ and $H=\dot{a}/a$. 
We focus on the flat case $K=0$. Combining eqs. \ref{ecmov} and working 
with the logarithmic time $p=\ln{a}+cte$ we have:
\begin{equation}
\frac{2}{(3-\phi^{\prime 2})} \phi^{\prime\prime}+ 
(1-\lambda)\phi^{\prime}=\frac{\sigma}{\rho}
\label{evolucion}
\end{equation}
where $\lambda=P/\rho$.

In Ref.\cite{DP94} it is analyzed both the radiation and the matter
dominated eras, with no cosmological constant. Even thought most of
the data, belong to the $\Lambda$ dominated era, we assume that the
expressions for the evolution of the dilaton field will be similar.
Therefore, we concentrate on the matter era, leaving the case with
cosmological constant to another work.  After decoupling, there is no
interaction between matter and radiation cosmologically important, so
we have $\rho=\rho_m+\rho_r \;$, $P=P_r+P_m \;$,
$\sigma=-\alpha_m(\phi)(\rho_m-3P_m)$ with $\; P_r=\frac{1}{3}\rho_r$,
$P_m\simeq 0$ and $\;
\alpha_m(\phi)=\frac{\partial\ln{m_m(\phi)}}{\partial \phi}$.\\ The
noninteracting relativistic and non relativistic components satisfy
respectively $\rho_r \propto a^{-4}$ and $\rho_m \propto
m_m(\phi)a^{-3}$. The evolution equation can be finally written as:
\begin{equation}
\frac{2}{(3-\phi^{\prime 2})} \phi^{\prime\prime}+ 
(1-\lambda(p,\phi))\phi^{\prime}=-[1-3\lambda(p,\phi)]\alpha_m(\phi)
\label{ecu1}
\end{equation}
where
\begin{equation}
3\lambda(p,\phi)=[1+C m_m(\phi) e^{p}]^{-1}
\label{ecu2}
\end{equation}
In the analysis of \cite{DP94} it was concluded that the radiation era
is very efficient in atracting the dilaton very near its minimum
$\phi_m$, and so we can approximate $m_m(\phi)\simeq const$ in
$\lambda(p,\phi)$.  We choose the origin of $p$ at matter-radiation
equality $\rho_r(p=0)=\rho_m(p=0)$, $\lambda(0)=1/3
\rho_r(0)/(\rho_r(0)+\rho_m(0))=1/6$.  Consequently $C=m_m^{-1}$ and
$\lambda(p)=\frac{1}{3}(1+e^p)^{-1}$.

From Eqs.(\ref{massexp},\ref{lnB}) we obtain
\begin{equation}
\alpha_A(\phi)=\beta_A(\phi-\phi_m)
\end{equation}
where  
\begin{equation}
\beta_A=\kappa (40.75- \ln(\frac{m_A}{1Gev}))
\label{beta} 
\end{equation}
and we have used the values as in Ref.\cite{DP94}. Even though we will
take for $m_A=1$ GeV, the mass of the nucleon, the dark matter case
$m=40 $ GeV would only change logarithmically the value of
$\beta$. Neglecting $\phi^{\prime 2}$ in Eq.\ref{ecu1} we obtain:
\begin{equation}
\frac{2}{3} \phi^{\prime\prime}+ 
(1-\lambda(p))\phi^{\prime}=-[1-3\lambda(p)]\beta_m(\phi-\phi_m)
\label{ecu3}
\end{equation}
Denoting $x=e^p=a/a_{equivalence}$ we have
\begin{equation}
(1+x)x \partial^2_x \phi+(\frac{5}{2}x+2) \partial_x 
\phi+\frac{3}{2}\beta_m(\phi-\phi_m)=0. \label{ecu4}
\end{equation}
This is a hypergeometric-type differential equation, with the
condition of regularity $\phi(x=0)=\phi_{rad}$, being $\phi_{rad}$ the
value of $\phi$ at the end of the radiation era. The parameters are
\begin{displaymath}
z=-x
\end{displaymath}
\begin{displaymath}
\alpha=\frac{3}{4}-iw
\end{displaymath}
\begin{displaymath}
\beta=\frac{3}{4}+iw
\end{displaymath}
\begin{displaymath}
\gamma=2
\end{displaymath}
with $w=\frac{3}{2}(\beta_m-3/8)^{1/2}$. The solution then reads
\begin{equation}
\phi-\phi_m = \left(\phi_{rad}-\phi_m\right) F(\alpha,\beta,\gamma;-x)
\end{equation}
with 
\begin{equation}
F_m(\alpha,\beta,\gamma;-x)=\frac{\phi(x)-\phi_m}{\phi_{rad}-\phi_m}
\end{equation}
As the argument is the red-shift, for high values we can use the 
asymptothic behaviour of $F_m$. Unless $\kappa$ has the unnatural value 
$9.2\times10^{-3}$ or less, the sign of $\beta_m-3/8$ will be 
positive, which we will assume. Then $\omega$ is real and we have:
\begin{equation}
F_m(\kappa,t)=\left[\frac{\coth(2 \pi \omega)}{\pi \omega 
(\omega^2+\frac{1}{16})}\right]^{1/2}\exp(-\frac{3}{4}p) \cos \theta 
\label{FM}
\end{equation}
where $\theta(t)= \omega p(t) +2 \omega \ln 2 +Arg \Gamma_2$ with 
$\Gamma_2=\Gamma(2i \omega)/\Gamma(2i \omega +\frac{3}{2})$ and 
$p=\ln(a(t)/a_{eq})$.
Defining  
\begin{equation}
F_r(\kappa)=\frac{ \phi_{rad}-\phi_m}{\phi_{i}-\phi_m}
\end{equation}
the atraction factor during the radiation era, whose value is estimated
to be $1.87\times10^{-4}\times\kappa^{-9/4}$ (see Ref\cite{DP94})
and
\begin{equation}
\triangle \phi=\phi_{i}-\phi_m
\end{equation}
with $\phi_i$ the ``initial'' (at some time $t_i$) value of the dilaton, 
we can write the solution for $\phi$ as:
\begin{equation}
\phi(t)-\phi_m = F_m(\kappa,t)F_r(\kappa)\triangle \phi
\end{equation}
or more explicitly:
\begin{equation}
\phi(t)-\phi_m =1.87 \times10^{-4}\times \kappa^{-9/4}
\left[\frac{\coth(2 \pi \omega)}{\pi \omega 
(\omega^2+\frac{1}{16})}\right]^{1/2}\left(\frac{a(t)}{a_{eq}}\right)^{-\frac{3}{4}} 
\cos \theta(t) 
\triangle \phi
\end{equation}\\

\subsection{Observational Adjustment}

Although it is tempting to work up to first or second order while analyzing 
Eq.\ref{delta1}, this cannot be done as we consider data spanning times 
comparable to the Hubble time. This makes the Taylor expansion of the 
function $\phi(t)$ meaningless. We will be forced then to make a non linear 
adjustment as we will see below.

We replace the solution found in the previous section into Eq.\ref{delta1}, 
obtaining (our convention for $\Delta$ is $\Delta f(t) \equiv f(t) - f(t_0)$) 
\begin{equation}
\frac{\Delta \alpha}{\alpha}(t) = 
1.75\times 10^{-8}{\cal A}\kappa^{-7/2}(\Delta\phi)^2 {\cal C}(\kappa) 
\Delta{\cal D}(\kappa,t), \label{delta2}
\end{equation}
where we have defined the functions
\begin{equation}
{\cal C}(\kappa)= \frac{\coth(2\pi\omega)}{\pi\omega(\omega^2+\frac{1}{16})}
\end{equation}
and
\begin{equation}
{\cal D}(\kappa,t)= \left(\frac{a(t)}{a_{eq}}\right)^{-3/2}
\cos^2\left(\theta(t)\right). 
\end{equation}
This entails a nonlinear adjustment of the parameters 
$\kappa^{-7/2}(\Delta\phi)^2$ and $\kappa$. 

Finally, it is important to note that not all data were fitted to
eq. \ref{delta2}. Actually the data obtained from experiments with
atomic clocks provide stringent bounds on $\frac{\dot
\alpha}{\alpha}$. Therefore, if we fit the derived quantity
$\frac{\Delta \alpha}{\alpha}$, the fitting procedure and results will
be almost dominated by these data. Thus we consider the following
expression for the actual data on the variation of $\alpha$
\cite{DP94}:

\begin{equation}
\frac{\dot \alpha}{\alpha}= - \kappa H_0 \left(\omega \tan
\theta(t_0)+ \frac34 \right) \left[1.87 \times 10^{-4} \kappa^{-9/4}
  F_m\left(\kappa,t_0\right) \Delta \phi \right]^2 
\label{dotalfa}
\end{equation}

\section{Results and Discussion}
\label{results}

\begin{figure}
\begin{center}
\includegraphics[scale=0.4,angle=-90]{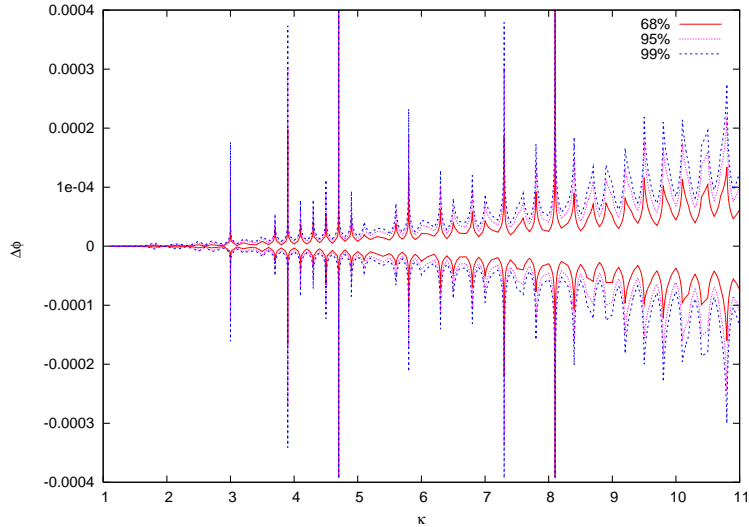}
\caption{Confidence contours for the free parameters of the DP dilaton model}
\label{contor}
\end{center}
\end{figure}

We have performed a statistical analysis working on a $\chi ^2$
function to compute the best-fit parameter values and uncertainties.
For the 246 data described in section \ref{bounds}, we obtain
$\chi^2_{min} = 357$. However, we have analized the contribution of
each data to $\chi^2$ and found that only $6$ data have an enourmous
contribution while the contributions of the other $240$ are of order 1
or less. Furthermore, excluding these data, we obtain $\chi^2_{min} =
240$ for 240 data and 2 free parameters with no change in the value of
the best fit parameters and errors. Therefore, we excluded these data
from the statistical analysis, using what we call the reduced data
set.  The contours of the likelihood functions in regions of 68 \%, 95
\% and 99 \% of confidence level are shown in figure
\ref{contor}. Unfortunately, the contours do not close, even if we
increase the values of $\kappa$. This is due to the fact that for
increasing $\kappa$, $\Delta \phi$ has to be increased in order to fit
the reduced data set (see eqs. \ref{delta2}, \ref{dotalfa}). However from figure
\ref{contor} we can find the following relation between the values of
$\kappa$ and $\Delta \phi$, that provide a good fit to the reduced
data set.

\begin{equation}
-3.4 \times 10^{-6} \kappa < \Delta \phi < 3.4 \times 10^{-6} \kappa
\end{equation}

On the other hand, table \ref{resultados1} shows constraints on the
values of $\Delta \phi$ for fixed values of $\kappa$. Again, the
values show how increasing values of $\kappa$, allow increasing
values of $\Delta \phi$ in order to fit the reduced data set.

We conclude that the string dilaton model proposed by Damour and
Polyakov is able to fit almost all experimental and observational data
on time variation of the fine structure constant. However, the
constraints obtained on the free parameters (considering one or two
free parameters) show that $\Delta \phi =0$ is also consistent with
the data and in this case the model predicts no variation of gauge
coupling constants or masses of elementary particles. Nevertheless, it
should be noted that in any grand unified theory the variation of the
fine structure constant is conected to the variation of the strong
coupling constant. Thus, the prediction of this model should also be
checked toghether with the bounds on the strong coupling constant
\cite{Ivanchik02,Ivanchik03}. Furthermore, the data considered in
section \ref{bounds} should be reanalized in order to relate the
observational values with variations in both the fine structure
constant and the strong coupling constant. This requires a more
detailed analysis of the QCD model considered and therefore we leave
it for future work.

\begin{table}[h]
\caption{The table shows for fixed values of $\kappa$, the best fit
  value and the corresponding $1 \sigma$ error of  $\Delta \phi$ } 
\label{resultados1}
\begin{center}
\begin{tabular}{|c|c|c|}
\hline $\kappa$ & $\Delta \phi$ & $ \sigma(\Delta \phi)$\\ 
\hline 
$0.01$& $2.4 \times  10^{-14}$&$2.3 \times 10^{-13}$\\\hline
$0.1$&$ -1.0 \times 10^{-10}$ &$ 8.0 \times 10^{-10}$\\ \hline
$1$&$-4.2 \times 10^{-8}$& $3.2 \times 10^{-7}$\\\hline
$10$&$3.6 \times 10^{-6}$&$2.7 \times 10^{-5}$\\\hline
\end{tabular}
\end{center}
\end{table}

\section{Acknowledgements}

S. Landau wants to thank  M. Murphy for early release of data and useful comments and A.Ivanchik for useful comments. 

\bibliography{landaubibliografia}
\bibliographystyle{aip} 
\end{document}